\documentstyle[aps,epsfig,prb]{revtex}

\begin{document}
\twocolumn
[\hsize\textwidth\columnwidth\hsize\csname@twocolumnfalse\endcsname

\title{Phase diagram of the half--filled Hubbard chain with
  next--nearest--neighbor hopping}
\author{ S.\ Daul $^{1,2}$ and R.M.\ Noack $^2$ }
\address{ $^1$ Physics Department, University of California, \\
                  Santa-Barbara CA 93106-9530. \\
$^2$ Institut de Physique Th\'{e}orique, Universit\'{e} de Fribourg, \\
   CH-1700 Fribourg, Switzerland.
}
\maketitle

\begin{abstract}

We investigate the ground--state phase diagram of the half--filled
one--dimensional Hubbard model with next--nearest--neighbor hopping
using the Density-Matrix Renormalization Group 
technique as well as an unrestricted Hartree--Fock approximation.
We find commensurate and incommensurate disordered magnetic
insulating phases and a spin--gapped metallic phase in addition to the
one--dimensional Heisenberg phase.
At large on--site Coulomb repulsion $U$, we make contact with the
phase diagram of the frustrated Heisenberg chain,
which has spin--gapped phases for sufficiently large frustration.
For weak $U$, sufficiently large next--nearest--neighbor hopping
$t_2$ leads to a band structure with four Fermi points rather than two,
producing a spin--gapped metallic phase.
As $U$ is increased in this regime, the system undergoes a
Mott--Hubbard transition to a frustrated antiferromagnetic insulator.

\end{abstract}
]


The one--dimensional Hubbard model is the prototypical model for
strongly interacting electrons in one dimension.
For repulsive interaction, its low--energy, long--distance physics is
well--described by the Luttinger liquid picture, in which the fundamental
excitations are gapless bosonic spin and charge modes, and the correlation
functions exhibit critical behavior with non--universal exponents
\cite{Luttingerliquid}.
At half filling, Umklapp processes lead to a gap in the charge
excitation spectrum and thus insulating behavior for
any finite value of the on--site Coulomb interaction, $U$.
The spin excitations behave as in the
strong--coupling, Heisenberg limit, i.e. are gapless with linear
dispersion. 

The introduction of a next--nearest--neighbor hopping can
change this picture dramatically.
In strong coupling, the additional hopping leads to a
frustrating next--nearest--neighbor Heisenberg interaction so that
the model maps to the frustrated Heisenberg chain.
At weak--coupling, the effect of $t_2$ is to change the band
structure, and, in particular, the number of Fermi points.
In this paper, we shall explore the interplay between the frustration at
strong coupling and the changed band structure at weak coupling.
As we shall see, the resulting phase diagram contains a number of highly
interesting phases: a spin--gapped metallic phase, 
commensurate and
incommensurate disordered magnetic insulating phases, as well as the
one--dimensional Heisenberg insulator.

\begin{figure}[htb]
 \begin{center}
  \epsfig{file=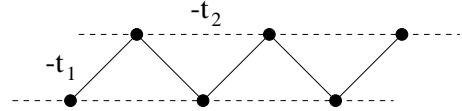,width=6cm}
 \end{center}
 \caption{The $t_1$--$t_2$ Hubbard chain.}
 \label{model}
\end{figure}

We study the Hamiltonian
\begin{eqnarray}
  \lefteqn{ H = -t_1 \sum_{i,\sigma} \left( c^{\dagger}_{i+1\sigma}
                                            c_{i\sigma} + h.c.  
 \right) } \nonumber \\
&&      -t_2\sum_{i,\sigma}\left( c^{\dagger}_{i+2\sigma} c_{i\sigma} 
                                  + h.c. \right)
    + U \sum_i n_{i\uparrow}n_{i\downarrow} \; ,
\label{eqhamt1t2}
\end{eqnarray}
where $t_1$ is the nearest--neighbor and $t_2$ the next-nearest 
neighbor hopping and $U$ is the on-site Coulomb repulsion.
It is useful to visualize the geometry as a zigzag structure as
depicted in Fig.\ \ref{model}.
\noindent
Here the summation goes over $L$ sites and spin $\sigma$, 
and we will always take $U$ positive and set $t_1=1$.
Since the sign of $t_2$ is irrelevant at half filling due to particle--hole
symmetry, we only consider $t_2>0$ in the following.
For $U=0$ and periodic boundary conditions, $H$ can be diagonalized
via a Fourier transform, yielding
\begin{equation}
  H = \sum_{k,\sigma} \varepsilon (k) c_{k\sigma}^{\dagger}c_{k\sigma} \; ,
\end{equation}
with $k$ an integer multiple of $\frac{2\pi}{L}$ and 
\begin{equation}
  \varepsilon (k) = -2t_1\cos k -2t_2\cos 2k \;.
\end{equation}
An interesting feature of this band structure is that there is a
nontrivial transition as a function of $t_2$ even at $U=0$.
For $t_2 < 0.5$, the noninteracting band has two Fermi points and
for $t_2 > 0.5 $, it has four Fermi points ($\pm k_F^{(1)}$ 
and $\pm k_F^{(2)}$).
This is important in a weak--coupling picture because the Fermi
points are separated by the Umklapp vector $q=\pi$ only for 
$t_2 < 0.5$.

For large $U$, Eq.\ (\ref{eqhamt1t2}) can be expanded perturbatively in
$1/U$, leading to the one--dimensional frustrated
Heisenberg Hamiltonian
\begin{equation}
  H = \sum_i \left( J_1 S_iS_{i+1} + J_2 S_iS_{i+2}  \right)
\end{equation}
with $J_1 = \frac{4t_1^2}{U}$ and $J_2 = \frac{4t_2^2}{U}$.
This model has been extensively studied using a number of different
methods \cite{WhiteAffleck}.
In particular, there is an exact solution at $J_2/J_1 = 0.5 $
($t_2/t_1 = 1/\sqrt{2}$) due to Majumdar and Ghosh \cite{MajumdarGhosh}.
At this point, the ground state is a simple dimer configuration and
the spectrum has a spin gap \cite{Affleck88}.
For $J_2/J_1 < J_c \approx 0.241167 $, studies combining numerical 
and field theory calculations \cite{Eggert96} have shown that the spin
excitation spectrum is gapless.
In the strong--coupling expansion,
this parameter value maps to the point at which the Fermi surface
jumps from two points to four in the $t_1$--$t_2$ Hubbard chain.
When $J_c < J_2/J_1 \le 0.5$, dimerization correlations are present and
for $J_2/J_1 > 0.5$, i.e. above the Majumdar--Ghosh point
\cite{MajumdarGhosh}, incommensurate spiral spin correlations appear.

\begin{figure}[htb]
\begin{center}
 \epsfig{file=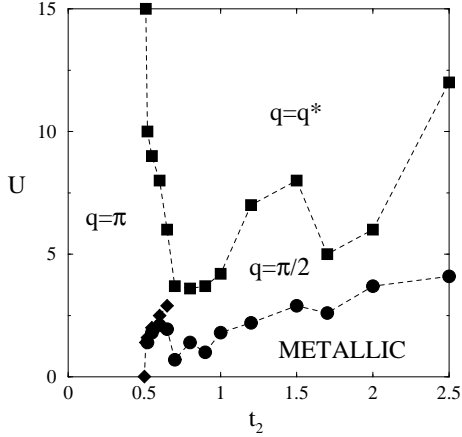,width=6cm}
\end{center}
\caption{The unrestricted Hartree-Fock phase diagram. The insulating phases
are labelled by the wave vector $q$ of the magnetic ordering for the
phases in which the order parameter $m>0$. }
\label{HFPhaseDiag}
\end{figure}

Mean--field theory can be a useful tool to provide qualitative
information about the ground--state phase diagram even though it 
tends to overemphasize ordered phases and cannot yield the correct
critical behavior.
Here we perform mean--field calculations starting from
the unrestricted Hartree-Fock Hamiltonian
\begin{equation}
 H_{\mbox{\footnotesize HF}} = 
H_0 + \frac{U}{2}\sum_\ell \left( \rho_\ell n_\ell - 
         {\bf m}_\ell \cdot {\bf s}_\ell  \right) 
\end{equation}
where $H_0$ is the non-interacting Hamiltonian, $n_\ell$ is the total
electron density on a site, 
and ${\bf s}_\ell = \sum_{s,s'} c_{\ell s}^\dag {\bf\sigma}_{s,s'}
c_{\ell s'}$ 
with $\sigma^{(x,y,z)}_{s,s'}$ the Pauli matrices.
The mean fields $ \rho_\ell$ and ${\bf m}_\ell $ are determined by the
self-consistent equations
$\rho_\ell = \langle n_\ell \rangle$ and 
${\bf m}_\ell = \langle {\bf s}_\ell  \rangle$.
We postulate a uniform density $\rho_\ell = \rho$, so that 
$\sum_\ell \rho_\ell n_\ell = \rho N$ is constant, and a spiral
arrangement of the magnetic moment
\begin{equation}
     {\bf m}_\ell = m \left(  \cos q\ell, \sin q\ell, 0  \right).
\end{equation}
The ground--state energy $E_0$ is then a function of the variational 
parameters $q$ and $m$ and the mean-field gap  is $\Delta = \frac{Um}{2}$. 
In order to obtain $q$ and $m$, we minimize $E_0(q,m)$ numerically for
large systems.
When $m=0$, the system is metallic, but
if $m$ is finite, the half--filled system is a magnetically ordered
insulator with principal wave vector $q$. 

We obtain the mean--field phase diagram shown in Fig.\ \ref{HFPhaseDiag}.
For $t_2<0.5$, the system is an antiferromagnet (i.e. $q=\pi$) for
all $U>0$.
This phase is the same as that obtained for $t_2=0$: 
The Fermi points are separated by the wave vector
$q=\pi$, leading to antiferromagnetic ordering.
For $t_2>0.5$, this commensurate separation of the Fermi points is
absent, and the system is a paramagnetic metal for small $U$.

As $U$ is increased, the system undergoes a transition to an
insulating phase with magnetic ordering at wave vector 
$q=k_F^{(2)} -k_F^{(1)}= \pi/2$. 
For still larger $U$, there is an incommensurate phase
with ordering at wave vector $q^\ast$, where 
$q^\ast$ goes continuously from $\pi$ at smaller $t_2$ to $\pi/2$ as
$t_2\rightarrow \infty$. 
As we shall see in the following, the phase boundaries and wave
vectors found in this mean--field phase diagram are qualitatively
similar to those found for the fully interacting system using the
DMRG, although the nature of the phases themselves is different.

\begin{figure}[htb]
\begin{center}
 \epsfig{file=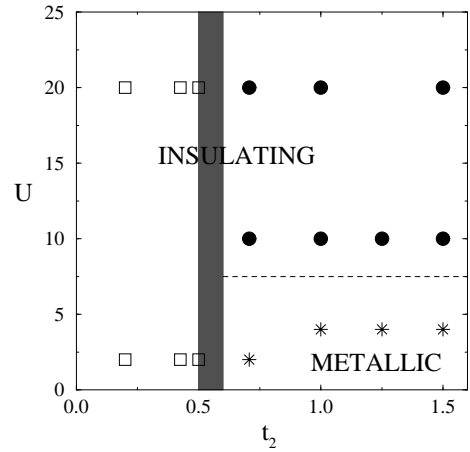,width=6cm}
\end{center}
\caption{DMRG phase diagram at half filling. 
  The symbol $\Box$ indicates a C0S1 phase, $\bullet$ a C0S0 phase
  and $\ast$ a C1S0 phase. 
  The approximate phase boundaries are marked as a guide to
  the eye.}
\label{DMRGPhaseDiag}
\end{figure}

Using the DMRG \cite{White92}, we can obtain the ground--state energy
and wave function on a finite lattice numerically to very high accuracy.
Here we perform calculations keeping up to 800 states on lattices of up to 64
sites so that the maximum weight of the discarded density matrix
eigenvalues is $10^{-6}$. 
In order to determine the phase diagram, we first calculate
the charge and spin gaps, defined as
\begin{eqnarray}
  \Delta_\rho &=& \frac{1}{2} \left[ E_0(N+2,0) + E_0(N-2,0)
  -2E_0(N,0) \right] \\
  \Delta_\sigma &=& E_0(N,1) - E_0(N,0)   \; ,
\end{eqnarray}
where $E_0(N,S)$ is the ground--state energy for $N$ particles
and spin $S$. 
We calculate the excitation gaps for systems with different size $L$ and then
extrapolate to $L\rightarrow \infty$ 
using a quadratic polynomial in $\frac{1}{L}$.
We have calculated the spin gap as a function of $U$ in the large $U$
limit (e.g. at $U=100$) and find very good 
quantitative agreement with numerical results for the frustrated Heisenberg
chain from Ref.\ \cite{WhiteAffleck}.

In Fig.\ \ref{DMRGPhaseDiag}, we display the phase diagram in the
$U$--$t_2$ plane determined using the $L \rightarrow \infty$ spin and
charge gaps computed with the DMRG.
We label the phases using the notation of 
Ref.\ \onlinecite{BalentsFischer}, in which C$n$S$m$ represents a
phase with $n$ gapless charge modes and $m$ gapless spin modes. 
For $t_2<0.5$, the system is in the one--dimensional Heisenberg phase
found for $t_2=0$, a C0S1 phase.
As in the mean--field phase diagram, this phase extends to the entire
region in which there are two Fermi points.
For $t_2>0.5$, there are two different regions both with gapped spin modes. 
For weak $U$, the system is metallic (C1S0),
and for $U > U_c$ of the order of the bandwidth, the
system is  insulating (C0S0).
In this region, where the system has four Fermi points, 
a weak-coupling renormalization group calculation predicts a metallic
spin liquid i.e. C1S0, phase,\cite{Fabrizio96}.
Such a spin--gapped phase has been found numerically for this model
away from half filling \cite{DaulNoack98}.
At half filling, there are no relevant Umklapp 
processes which could drive the system to an insulating phase when
there are four Fermi points \cite{Fabrizio96}.

\begin{figure}[htb]
\begin{center}
 \epsfig{file=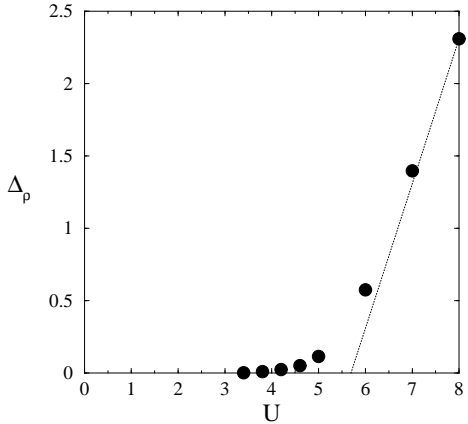,width=6cm}
\end{center}
\caption{Charge gap as a function of $U$ for $t_2=1$ at half filling.
The dotted line is linear in $U$. 
}
\label{ChargeGap}
\end{figure}

In Fig.\ \ref{ChargeGap}, we show the charge gap as a function of $U$ for 
$t_2=1$. 
There is a clearly defined metal--insulator transition at 
$U_c = 3.2 \approx W/2$, where $W$ is the bandwidth of the
non-interacting case ($W=6.25$ for $t_2=1$).
For $U \gg U_c$, the gap grows linearly as $U-U_c$ as one would expect
above a Mott--Hubbard transition \cite{motthubbard}, but the curve is
rounded near the transition point.
The metal--insulator transition as a function of $U$ therefore appears to be 
continuous. 

\begin{figure}[htb]
\begin{center}
 \epsfig{file=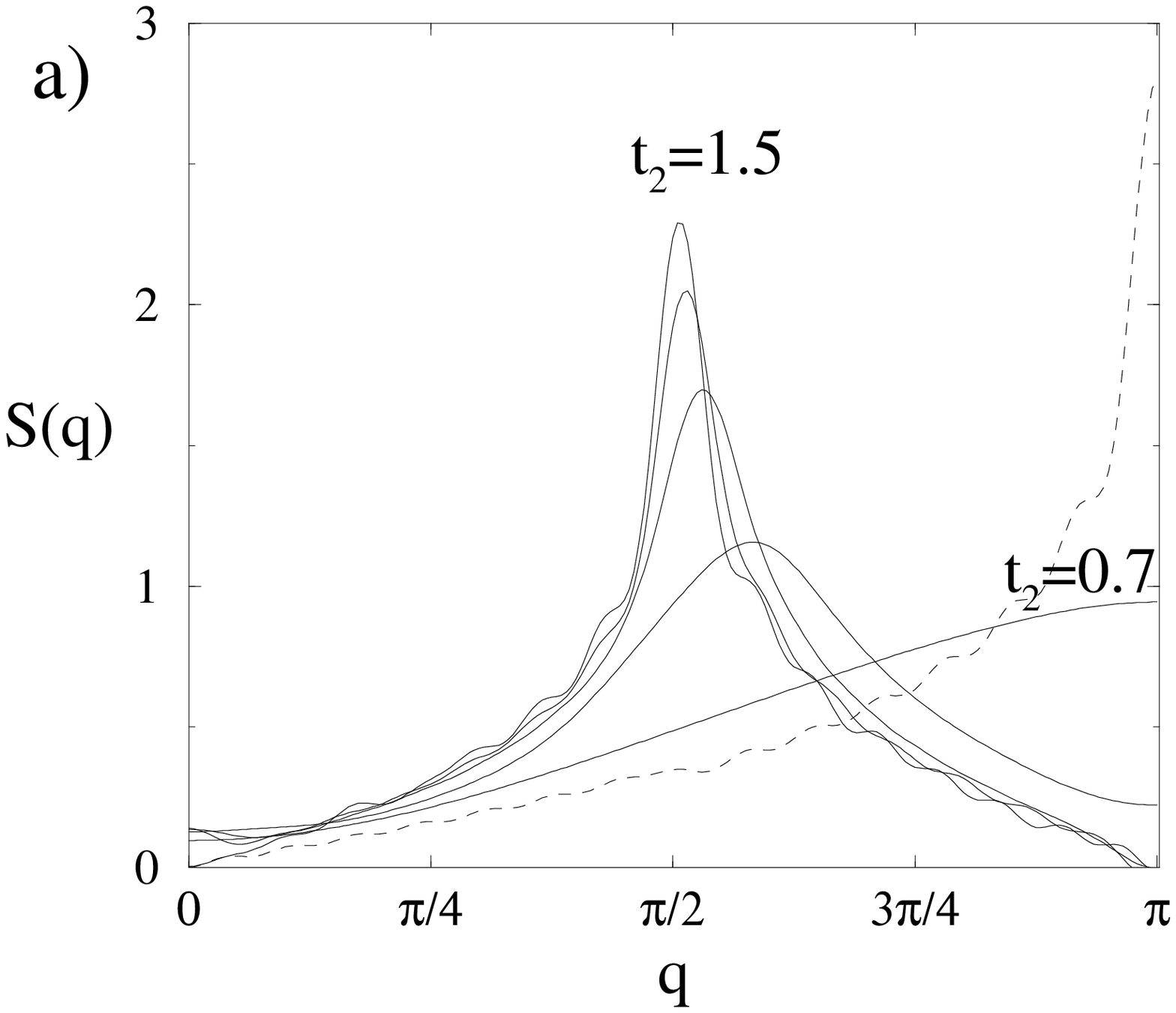,width=6cm}
 \epsfig{file=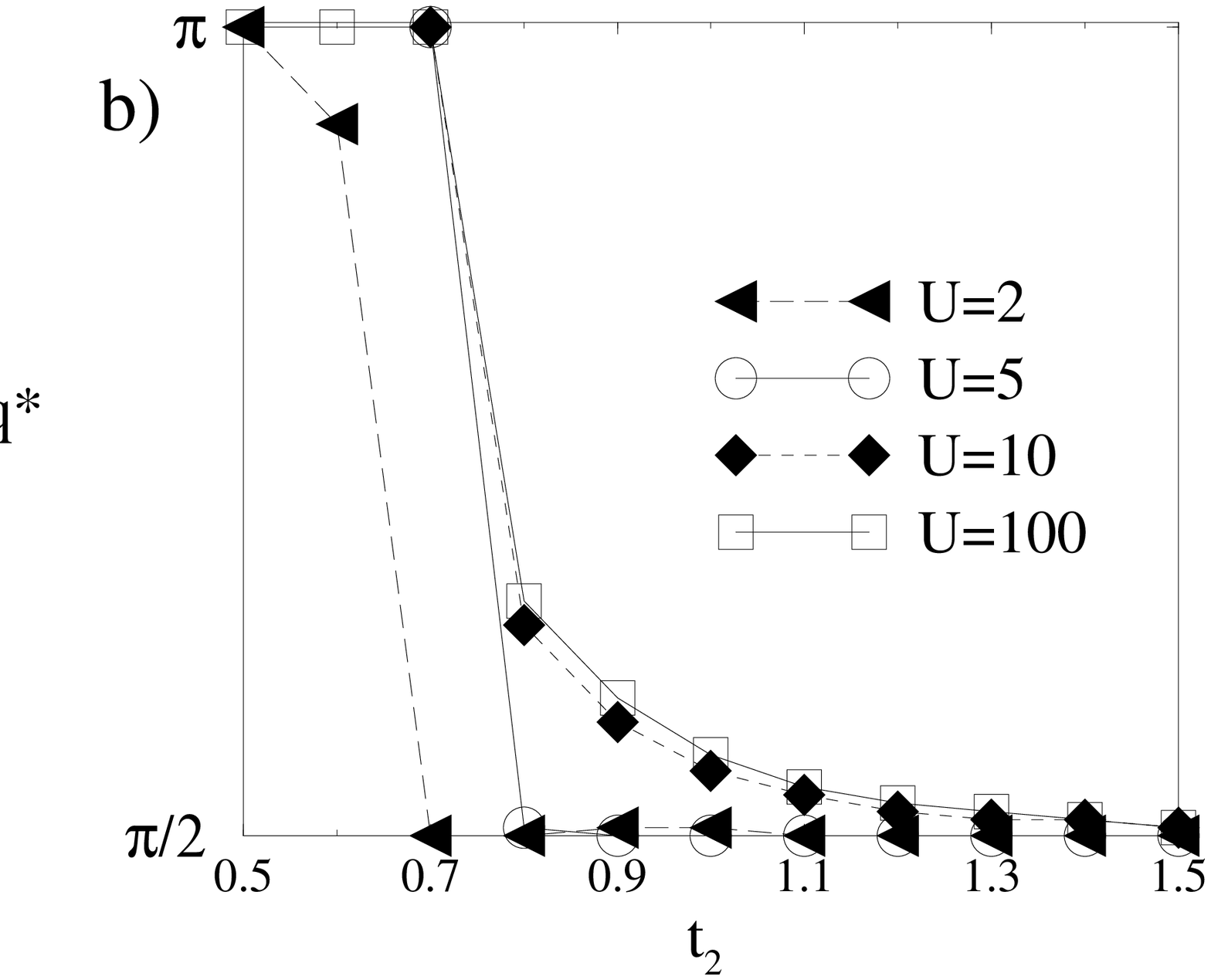,width=6cm}
\end{center}
\caption{(a) Fourier transform, $S(q)$, of the spin--spin correlation
function for the half--filled system with $L=40$, $U=100$ and 
$t_2 = 0.7 \ldots 1.5$ in steps of 0.2 (solid lines from bottom to top at
$q=\pi/2$); the dashed line is the result for $t_2=0$. (b) The wavevector 
$q^\ast$ obtained from the position of the peak in $S(q)$ as a 
function of $t_2$ for various values of $U$.}
\label{SpinCorrelationFunction}
\end{figure}

We now investigate the nature of the magnetic ordering in the different phases
by examining the spin--spin correlation function
\begin{equation}
 S_{\rm av}(r) = \sum_{\{\ell\}}\langle s_\ell^+s_{\ell+r}^-\rangle
\end{equation}
where $s_\ell^+$ ($s_\ell^-$) are the spin raising (lowering) operators
corresponding to ${\bf s}_\ell$, and 
we average over a number of $\ell$--values (typically six) to
reduce oscillations present because of the open boundaries.
We then perform a continuous Fourier transform to obtain the static
structure factor
\begin{equation}
   S(q) = \int_{-\infty}^\infty 
   dr \; e^{iqr} S_{\rm av}(r) \; .
\end{equation}
The resulting function is plotted in
Fig.\ \ref{SpinCorrelationFunction}(a) for various $t_2$ ranging from
0.7 to 1.5.
It is clear that the peak shifts continuously from $q=\pi$ at $t_2=0.7$
to $q=\pi/2$ at $t_2=1.5$. 
As $t_2$ increases, the peak at $q= \pi/2$ becomes sharper, resembling
more closely the $q=\pi$ peak of $t_2=0$ structure factor, also shown
for reference.
This is because the correlation length diverges as $t_2$ becomes
large \cite{WhiteAffleck}; it is the limit of two
uncoupled Heisenberg chains.

The position of the peak, $q^\ast$, is plotted as a function of $t_2$ in 
Fig.\ \ref{SpinCorrelationFunction}(b) for different $U$ values.
For $U\geq 5$, 
$q^\ast=\pi$ for $t_2 < t_2^\ast \approx 0.7$.
At large $U$, one expects $q^\ast$ to deviate from $\pi$ only for $t_2$
above the Majumdar--Ghosh point, 
$t_2^{\rm MG}=1/\sqrt{2}$ \cite{WhiteAffleck}.
As can be seen, $t_2^\ast \approx t_2^{\rm MG}$ down to small values
of $U$. 
Note that $t_2^\ast$ does not coincide with the opening of the spin gap in 
Fig.\ \ref{DMRGPhaseDiag}; there is an  
intermediate region of a spin--gapped dimer phase with $q^\ast=\pi$.
As $U$ is reduced, the region of incommensurate
spiral spin order becomes narrower. 
This is expected because there is no incommensurate spiral spin 
order at $U=0$. 

\begin{figure}[htb]
\begin{center}
 \epsfig{file=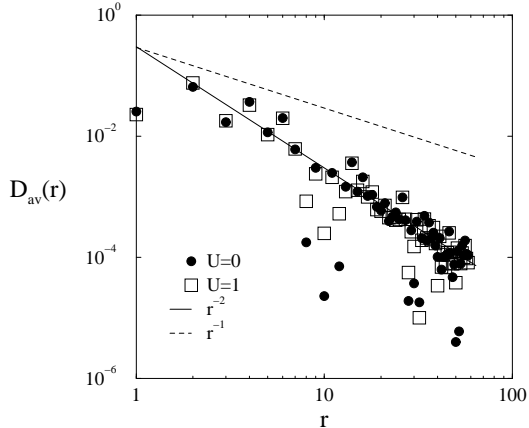,width=6cm}
\end{center}
\caption{Pairing correlation function on a log-log scale for $L=64$, 
    $t_2=0.78$ and $U=0,1$. } 
\label{PairingCorrelationFunction}
\end{figure}

In the metallic spin--liquid (C1S0) region, the long--distance behavior is
characterized by the correlation functions associated with the
charge degrees of freedom.
A bosonization treatment valid for weak $U$ \cite{Fabrizio96} predicts
that two competing correlation functions have the slowest asymptotic decay: 
the dimer--wave correlation function
\begin{equation}
 \chi_{DW} (x)   \sim \frac{\cos \pi x}{x^\theta}
\end{equation}
and the singlet superconducting  correlation function 
\begin{equation}
  \chi_{SC}(x)   \sim \frac{1}{x^{1/\theta}} \; .
\end{equation}
If $\theta>1$, the pairing is dominant, 
while for $\theta<1$ the dimer wave is dominant. 
In addition, at half filling high-order Umklapp process are relevant
for $\theta < 1$ and the system becomes insulating \cite{Fabrizio96}.
We then expect that $d$-wave pairing correlation are dominant for
$U < U_c$, where the charge gap opens at $U_c$.
We compare the pairing function for the weakly interacting and 
the non-interacting case to see if there is an enhancement.
At half filling $k_F^{(2)} - k_F^{(1)} = \pi / 2$, so the relevant pair 
operators in real space involve pairs on next--nearest--neighbor sites. 
We have carefully evaluated the pairing correlation function  
\begin{equation}
    D_{\rm av}(r) = \sum_{\{i\}} \langle \Delta_i \Delta_{i+r}^+ \rangle
\end{equation}
where 
$\Delta_i = (c_{i\uparrow}c_{i+2\downarrow} 
- c_{i\downarrow}c_{i+2\uparrow}) / \sqrt{2} $
and we average over a number of $i$--values to
reduce oscillations present because of the open boundaries.
In Fig.\ \ref{PairingCorrelationFunction}, we plot the results for $U=0$ and 
$U=1$, and we clearly see that there is no enhancement when we turn $U$ on.
This is in contradiction with previous work by Kuroki et al. \cite{Kuroki97} 
who claim that the superconducting correlations are dominant for small $U$ 
based on projector QMC. 
We have also evaluated the dimer-wave correlation function which shows no 
enhancement for $U < U_c$, but the long range behavior clearly changes 
from $r^{-2}$ to $r^{-1}$ when $U > U_c$.

In conclusion, we have explored the rich ground--state phase diagram of the
half--filled  $t_1$--$t_2$ Hubbard chain using the Density Matrix
Renormalization Group.
For $t_2 < 0.5 $ the model is insulating for 
all values of $U$ due to Umklapp processes and there is no spin gap. 
For $t_2 > 0.5$ the system has a spin gap for all $U$ but is metallic
(i.e. the charge gap vanishes) for sufficiently small $U$.
A continuous metal to insulator transition occurs when $U$ is of the order 
of the bandwidth.
At large $U$, we find quantitative agreement with the phase diagram of
the frustrated Heisenberg chain.
While the overall phase diagram is in agreement with weak--coupling
renormalization group calculations \cite{Fabrizio96}, the pairing 
correlations do not behave as predicted in the
spin--gapped metallic phase.
The exact nature of the metallic phase is difficult to determine;
more work must be done in this direction.

This work was supported by the Swiss National Foundation under Grant
Nos. 20--46918.96 and 20--53800.98.
We would like to thank D.\ Baeriswyl, F.\ Gebhard and D.J.\ Scalapino
for helpful discussions.


\end{document}